\documentstyle[twoside,fleqn,espcrc2,epsf]{article}

\newcommand{\AmS}{{\protect\the\textfont2
  A\kern-.1667em\lower.5ex\hbox{M}\kern-.125emS}}

\title{ Extended instantons generated on the lattice}

\author{E. Mendel\address{FB Physik,
 Carl von Ossietzky Universit\"at Oldenburg\\
         26111 Oldenburg, Germany}%
        \thanks{Dedicated to the memory of my beloved
brother,\ \ \ \ \ \ \ \ \ \ \ \ \ \ \ \ \ \ \ \ \ \  Prof. Roberto Mendel.}
        and 
   G. Nolte  \thanks{New Address: \  Abteilung f\"ur Neurobiologie,
  Universit\"atsklinikum Steglitz, 12200 Berlin.} }
       
\begin{document}

\begin{abstract}
   We have been able to observe directly extended
 instantons on the lattice, with a new method that does 
 not require dislocations to measure them, and where we
 do not perform cooling.   
    We showed, based on the simple Abelian Higgs model in 
 $1+1$ dim., that one can extract the instanton and 
 anti-instanton density and their size, by measuring the 
 topological charge, $Q_v$ , on sub-volumes $v$ larger than
 the instanton sizes, but smaller than the periodic lattice
 of size $V$. We are working on the generalization for 
 non-abelian models.
\end{abstract}

\maketitle

\section{Introduction}
  Instantons \cite{Bela,tHoo} are expected
to play an important role in such diverse phenomena as net Baryon number
generation, the $U(1)$ problem or even quark confinement. In order to
extract their Quantum contribution to the path integral it is desirable
to use a lattice formulation.  We will propose a method that enables
the observation of these topological non trivial configurations, even 
for relatively smooth configurations that don't present lattice artifacts
called dislocations. (The topological charge
is usually only measured on the whole volume $V$ where, as a consequence of
periodic boundary conditions (PBC), we can only see a net charge if we have 
dislocations).  Furthermore, we will show that one can extract information
on the instanton  density $\rho_I$ and size $L_I$, without the need of
cooling \cite{Hoek} to suppress the quantum fluctuations. Note that
one will still need improved actions \cite{Garc} or cooling, for
$4$ dimensional gauge theories, in order to avoid small instantons  
comparable with the lattice spacing $a$. If eventually ``perfect actions''
are found that would avoid dislocations, our method  could be very
useful to enable the extraction of topological information.  

 The  method \cite{Our}, which we present here, 
consists in
measuring the topological charge on sub-volumes $v$ (of $V$), larger than 
or of the order of
the relevant instantons (I, AI). Let us first assume that one can reach a  
coupling
regime where the gauge configurations are smooth enough so that there are
no individual sites where the topological density gets even close to $\pm 1/2$.
This can  be reached for the abelian Higgs model in $1+1$ dimensions
at weak coupling.
Then one has safely  configurations without dislocations and if we take
PBC the total geometric topological charge is 
exactly zero.  This just means that the charge for a number of I and AI
has to add to zero.

  We will use as our main observable, the  probability distribution to get 
some topological charge $Q$ in sub-volumes $v$,  $P_v (Q)$. 
 For very large $v$, this distribution 
will have lumps peaked at integer values, representing a net number of 
I minus AI in $v$.      
 As we reduce the volume $v$ to sizes smaller
than the instanton, we can only see a distribution peaked at $Q=0$  as we
have just pieces of instantons in $v$ or  quantum fluctuations.
 Interestingly, these are $Q$ fluctuations in regions of stable
topological vacua and therefore tend to cancel out over fairly short
distances. Due to their topological character
the total dispersion $\sigma$, that they produce in $Q$,
only grows as $\sqrt {s}$ with $s$ being the free surface.
 For larger volumes,
of the order or somewhat larger than the instantons, we start getting
relevant contributions to  $P_v (Q)$ from instantons, where the charge 
adds coherently to $\pm 1$ (if they are fully in $v$)
in contrast to the fluctuations. 

 The idea is then to look at  scales $v$, where: 
\vspace{-1mm}
$$
    a^D \ll v_I < \ \ v \ \ < V  
$$
\newpage \noindent
so that we have mainly suppressed the fluctuations, and can study the ensemble
of I and AI  with effectively almost free spatial BC.

 For the abelian Higgs model on fine lattices, we have found
clearly identifiable extended instantons over several lattice spacings,
while the charge over the whole $V$ stays always zero. Assuming
an almost free ensemble of I and AI, we have been able to extract
an  instanton average density $\rho_I$ and size $L_I$ (also as function of
temperature, $T=1/N_\tau$). The density $\rho_I$ which can also be
extracted from the second moment of  $P_v (Q)$ for large $v$,
falls steeply at higher $T$ as the
instantons get squeezed in the time direction. We have 
also studied \cite{Our}  the evolution of the position of the instantons 
in Monte Carlo time and found that they  get created locally as 
I - AI pairs and annihilate later when they encounter another pairing 
possibility.  

  As for some  models it is  hard to avoid dislocations, we have
analyzed for our case what happens if one takes rougher lattices
where dislocations start to appear. Only the net number $N$ of dislocations
matters (pairs can be gauged away), effectively behaving as modified
boundary conditions with $N$ instantons. We have found in this coupling
regime, that the
probability to be in a sector with $N$ net instantons in the full
volume gives a compatible
distribution to the case without dislocations in sub-volumes of
the same size.
  
  The previous discussion has been done using the geometrical
definition for the topological density \cite{Lusch,Pana} but
we have also investigated the goodness of the field theoretic definition
which is easier to implement \cite{DiVec} and
gives the right naive continuum limit, but does have only
approximately right addition properties and so does not give exactly 
integers on the full volume $V$.

\section{ Topological Q for Abelian Higgs model}

  The lattice formulation of the abelian Higgs model in 1+1 dim., which 
is one of the simplest allowing non-trivial topology, has the action $S_n$: 
\begin{equation}
\lambda\big(\Phi_n^\ast\Phi_n-1\big)^2-\beta Re(U_n)
        -2\kappa Re(\Phi_n^\ast U_{n,\mu}\Phi_{n+\mu})
\end{equation} 
 Here $\Phi_n$ is the Higgs field,  
$U_{n,\mu}= \exp(i\theta_{n,\mu}) $ are the links 
and $U_n= \Pi_{\Box}U_{n,\mu} $ the  $2$-d plaquettes. 

  In the continuum the topological charge in $v$ is
\begin{equation} 
Q_v=\frac{1}{4\pi}\int_v d^2x\ \epsilon_{\mu\nu}F_{\mu\nu}
   =\frac{1}{2\pi}\int_{\partial v} ds\  n_\mu A_\mu  \ . \label{bound} 
\end{equation} 
so that on the whole $V$  it is always zero for PBC.
   
  On the lattice the field-theoretic definition, giving the right
naive continuum limit,
 \begin{equation} 
Q^F_v :=\frac{1}{2\pi}\sum_{n\in v} Im(U_n) , \label{def.f}
\end{equation} 
has the disadvantage of not being a proper topological density,
not satisfying exactly the second equality in eq.(\ref{bound}) and so
it isn't strictly $0$ in $V$.

  The geometrical charge definition,
which is the simplest example of the fibre bundle
construction used in the non-abelian case \cite{Lusch,Pana}, is
\begin{equation} 
Q^G_v :=\frac{1}{2\pi i}\sum_{n\in v}\log(U_n). \label{def.g1} 
\end{equation} 
This $Q^G_v$ may still violate eq.(\ref{bound}) by an integer amount in  cases
where the phases in the links of a plaquette add up to more than $\pm \pi$,
this value being reduced by $\pm 2 \pi$, by the $\log$ to its principal
branch. These lattice artifacts, called dislocations, are the reason for
obtaining net charge on a periodic lattice.   These  artifacts
can be avoided on fine enough lattices at least for this model.
         
 In order to keep track of these dislocations  we will directly work 
with the phases $\theta_{n,\mu}$ of the links 
as the fundamental variables. For  the plaquettes,  
 $\theta_n=\theta_{n,\hat{x}}+\theta_{n+\hat{x},\hat{\tau}}
-\theta_{n+\hat{\tau},\hat{x}} -\theta_{n,\hat{\tau}}$. 
We then define
\begin{equation} 
Q_v:=\frac{1}{2\pi }\sum_{n\in v}[\theta_n] \ , \label{def.g2} 
\end{equation} 
where the brackets shift $\theta_n$ by an integer multiple of 
$2\pi$ into the interval $(-\pi,\pi]$.  
This definition is clearly equivalent to (\ref{def.g1}), however, using 
$\theta_n$ as the fundamental variable we are now able 
to locate  dislocations, namely places for which 
$|\theta_n| > \pi $.
Note that in our Metropolis algorithm we  update the $\theta_{n,\mu}$
by small changes from the old values, otherwise we could trivially 
generate dislocations by adding $2 \pi$ to a link. 
 We find  that if we choose parameters corresponding
to a fine enough lattice, so that the physical instantons are much larger
than $a$, we can avoid obtaining any dislocations at all over the whole
Monte Carlo run.  Most simulations have been done then around the 
``scaling region ''  \cite{Our}, $( \lambda=.2, \kappa=.37, \beta > 7.  )$.
\begin{figure}[bth]
\vspace{4mm}
\hspace{-8mm}
\epsfxsize=8.2cm
\epsffile{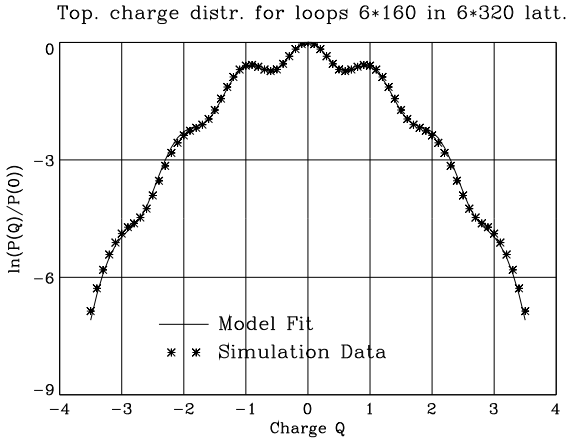}
\vspace{-10mm}
\caption{Probability  to find a  charge $Q$ in a volume 
$v$, half of $V$. Note the data agreement with
the model over $3$ orders of magnitude. The rise at integers
is due to instantons fully in $v$. }
\vspace{-4mm}
\end{figure}
\section{Identifying Instantons}

  In order to be able to identify instantons clearly, we have chosen lattices
with spatially very elongated $V$ and with subvolumes $v$
which wrap around the $\tau$ axis in order to suppress the charge fluctuations
along that long ``boundary''. The finite temperature also serves to get a 
diluter system of instantons. We took for example $V = 320 \times 6$ 
and $v = (20-160) \times 6$ and could clearly see the creation of 
I - AI pairs.
  
  The charge probability distribution $P_v(Q)$ is shown in fig. 1, with clear
peaks at integer charge. For a  quantitative analysis to extract $\rho_I$,
$L_I$ and the dispersion in the fluctuations $\sigma$, we developed a simple
model of almost noninteracting instantons. Assuming a number density
 $\rho_I = \rho_{AI}$, the probability to have $n$ pairs in the total $V$ is
\begin{equation}
p_n(\rho V)=\frac{(\rho V)^{2n}}{n!^2}\frac{1}{N} \ ,
\end{equation}   
where $N$ is the normalization. This $p_n$ has to be folded with the
number of ways to get a charge $Q$ in $v$, assuming that the  $n$ instantons
of size $L_I$ can be located anywhere in $V$. With this distribution
$f_n^v(Q)$ in a given sector, we get
\begin{equation}
P_v(Q) =\sum_n p_n f_n^v(Q) \ . 
\end{equation}   
This ideal distribution can be smeared
by a  gaussian with a  width $\sigma$ due to quantum fluctuations.
The best fit for the parameters in lattice units is   
$\rho \approx .0011$,  $L_I \approx28$ and $\sigma \approx .27$. The
topological susceptibility can also be modeled \cite{Our} for large
$V$ being $<Q^2_v> \approx 2 v \rho_I$. We have checked compatibility
for various $v$, and calculated at some T.

\section{Beyond optimal conditions}

  As for some models we cannot reach couplings where we can suppress
dislocations, it is interesting to compare
 for a $\beta=6.3$, where they just start
to appear, the $P_{Vd} (Q)$ for sectors with net number of dislocations
(and net instantons) with the $P_v (Q)$ without dislocations with $v=V_d$.
 The agreement is shown in Table 1, telling us that the net instantons come
with the right $S$ and entropy. In this sense the traditional way to measure
topology is fine, at least for large instantons.
 
  We also used  the field theoretic  $Q_v^F$ and found
good agreement \cite{Our} with $Q_v^G$  for fine lattices. For
$SU(2)$ we have observed much smaller fluctuations for $Q_v^F$ if,
instead of symmetrizing loops as in Ref.\cite{DiVec}, we reorder terms (same 
$Q$ on full $V$) taking for each  hypercube all loops wrapping it.
  
  \begin{table}[thb]
\vspace{-8mm} 
 \begin{center}
 \caption{Relative weights of sectors with charge $|n|$.} 
 \begin{tabular}{|c|c|c|c|c|}\hline
    n       & 0  & 1 & 2 & 3 \\ \hline 
 with disloc. & 1.  & .51(4) & .11(1.5) 
  & .005(2) \\ \hline    
 no disloc. & 1. &.53(2)   & .10(1) 
  & .007(3)  \\ \hline 
 \end{tabular} 
 \end{center}
\vspace{-11mm}  
 \end{table}
 

\begin{thebibliography}{00100}
\bibitem{Bela} A. Belavin and A. Polyakov JETP Lett. 22 (1975) 245.
\bibitem{tHoo} G. 't Hooft, Phys. Rev. Lett. 37 (1976) 8.
\bibitem{Hoek} J. Hoek, M. Teper and J. Waterhouse, Nucl. Phys. B288
  (1987) 589.
\bibitem{Garc} M. Garcia et al, Nucl. Phys.B34(Proc.Suppl.) (1994) 222;
 P. de Forcrand, M. Garcia and \ I. Stamatescu, hep-lat/9509064.
\bibitem{Our} E. Mendel and G. Nolte, hep-lat/9511030. 
\bibitem{Lusch} M. L\"uscher, Com. Math. Phys. 85 (1982) 39. 
\bibitem{Pana} C. Panagiotakopoulos, Nucl. Phys. B251 (1985) 61.
\bibitem{DiVec} P. diVecchia et al,  Nucl.Phys.B192(1981)392 
\end{thebibliography}
\end{document}